\documentclass[preprint,aps,prd,superscriptaddress,preprintnumbers,nofootinbib]{revtex4-1}

\usepackage[utf8]{inputenc}
\usepackage{graphicx}
\usepackage[T1]{fontenc}
\usepackage{amstext}
\usepackage{amsmath,amssymb,array}
\usepackage{babel}
\usepackage{color}
\usepackage{hyperref}

\begin{document}
	
	\title{$N_{\textrm{eff}}$ Constraint on Pseudo-Dirac Neutrinos}
	
	\author{Chee Sheng Fong}
	\email{sheng.fong@ufabc.edu.br}
    \affiliation{Centro de Ciências Naturais e Humanas,
Universidade Federal do ABC, 09.210-170, Santo André, SP, Brazil}
	\author{Yago Porto}
	\email{yago.porto@tum.de}
    \affiliation{Centro de Ciências Naturais e Humanas,
Universidade Federal do ABC, 09.210-170, Santo André, SP, Brazil}
\affiliation{Physik-Department, Technische Universit{\"a}t München, James-Franck-Stra{\ss}e, 85748 Garching, Germany}

	\begin{abstract}
		After the electroweak symmetry breaking, we can write down two types of mass for the Standard
		Model neutrinos, Dirac or Majorana. It is often said that both
		types of mass cannot be distinguished in neutrino oscillation phenomena.
		This is in fact not true if neutrinos are pseudo-Dirac (strictly speaking still Majorana) where they
		mix almost maximally with sterile neutrinos to form pseudo-Dirac pairs.
		If this is indeed realized in Nature, what we observe experimentally
		as three mass eigenstates are actually three pairs of mass eigenstates
		with yet-to-be-measured new mass splitting among each pair. While
		the new mass squared splitting of the first and second mass eigenstates
		have stringent constraints from solar neutrino to be $|\delta m_{1,2}^2| \lesssim10^{-11}\,\textrm{eV}^{2}$,
		the one regarding the third mass eigenstate has a weaker constraint $|\delta m_3^2| \lesssim10^{-5}\,\textrm{eV}^{2}$.
		By keeping only one nonzero pseudo-Dirac mass squared splitting at a time, we derive an effective 3+1 description for the pseudo-Dirac scenario.
		Then we use the Cosmic Microwave Background (CMB) constraint on neutrino relativistic degrees of freedom $N_{\textrm{eff}}$ to derive a
		new constraint $|\delta m_3^2| < 2 \times 10^{-6}\,{\rm eV}^2$ and show that the future CMB-S4 and CMB-HD can improve this bound by an order of magnitude.
	\end{abstract}

	\maketitle
	
	\newpage

\section{Introduction}

While neutrino oscillation experiments have shown that at least two
of three of the Standard Model (SM) neutrinos have nonzero masses,
the nature as well as origin of their masses remain an open problem.
It is often said that the nature of neutrino mass, whether Dirac or
Majorana, cannot be distinguished in neutrino oscillation experiments.
This is not true if neutrino masses are pseudo-Dirac where the SM
neutrinos mix almost maximally with sterile neutrinos to form pseudo-Dirac
pairs (strictly speaking, all of them have Majorana masses) \cite{Valle:1982yw,Wolfenstein:1981kw,Petcov:1982ya}.\footnote{The same scenario is referred to as quasi-Dirac in refs.~\cite{Anamiati:2017rxw,Anamiati:2019maf,Fong:2020smz}.} If three sterile neutrinos are introduced, we can have three pseudo-Dirac
pairs $(j,j+3)$, $j=1,2,3$ with masses
\begin{eqnarray}
\hat{{\cal M}}_{j}^{2} & \equiv & m_{j}^{2}-\frac{1}{2}\delta m_{j}^{2},\quad\hat{{\cal M}}_{j+3}^{2}\equiv m_{j}^{2}+\frac{1}{2}\delta m_{j}^{2}.
\end{eqnarray}
where $\delta m_{j}^{2}$ are the three new mass squared splitting
not yet measured in any experiment. The mixing between the SM and
the sterile neutrinos is close to maximal 
where the deviations from maximality are expected to be small due to the pseudo-Dirac structure. 

Under the assumption of maximal mixing, strong experimental constraints
using solar neutrino data have been derived $|\delta m_{1,2}^{2}|\lesssim10^{-11}\,\textrm{eV}^{2}$
while it is not sensitive to $\delta m_{3}^{2}$ due to small $\theta_{13}$~\cite{Cirelli:2004cz,deGouvea:2009fp,Anamiati:2017rxw,Ansarifard:2022kvy}.
Using atmospheric, beam and reactor neutrino data which are sensitive to
atmospheric mass splitting and $\theta_{13}$, a much weaker constraint
is obtained $|\delta m_{3}^{2}|\lesssim10^{-5}\,\textrm{eV}^{2}$~\cite{Anamiati:2017rxw}.
(See also \cite{Cirelli:2004cz}.) Due to the small $\delta m_j^2$,
astrophysical neutrinos coming from distance sources of Mpc to Gpc
allow to probe mass squared difference much smaller than $10^{-12}\,\textrm{eV}^{2}$
\cite{Keranen:2003xd,Beacom:2003eu,Cirelli:2004cz,Esmaili:2009fk,Esmaili:2012ac,Joshipura:2013yba,Brdar:2018tce,DeGouvea:2020ang,Martinez-Soler:2021unz,Rink:2022nvw,Carloni:2022cqz,Fong:2024msb,Dev:2024yrg}. In particular, using IceCube diffuse all-sky flux measurements, sets $3\sigma$ constraints on quasi-Dirac mass splittings in the range $(5\times10^{-19},\,8\times10^{-19})\,\mathrm{eV}^2$~\cite{Carloni:2025dhv}.

Cosmological probes like additional effective relativistic degrees
of freedom $N_{\textrm{eff}}$ during neutrino decoupling era and
the light element abundances produced from the Bang Big Nucleosynthesis
(BBN) can also be used to constrain pseudo-Dirac scenario. In refs.
\cite{Kirilova:1997sv,Kirilova:1999xj,Kirilova:2006wh}, a 1+1 (active
+ sterile) model together with the BBN measurement of $^{4}$He abundance
have been used to constrain to new mass squared splitting to be below
$10^{-8}\,\textrm{eV}^{2}$ for maximal active-sterile mixing while a more
realistic 3+1 model considered in refs.~\cite{Dolgov:2003sg,Cirelli:2004cz}
gave a less stringent bound $|\delta m_{3}^{2}|\lesssim10^{-5}\,\textrm{eV}^{2}$.
In comparison to the BBN determination of relativistic degrees of freedom in terms of neutrino species $N_{\rm eff}$~\cite{Schoneberg:2024ifp}, our aim in this work is to derive constraint
on $\delta m_{3}^{2}$ using the much more precisely measured $N_{\textrm{eff}}$ from the Cosmic Microwave Background (CMB).
Current CMB measurements from the Planck satellite combined with baryon acoustic oscillation (BAO) measurements from galaxy surveys give $N_{\textrm{eff}}=2.99^{+0.34}_{-0.33}$ at 95 \% CL~\cite{Planck:2018vyg} consistent with the most precise
theoretical prediction of the SM in the standard cosmology $N_{\textrm{eff}}=3.0440\pm0.0002$~\cite{Bennett:2020zkv} while the future ground-based telescopes CMB-S4 will be sensitive to $N_{\textrm{eff}}$ at the level of $0.06$~\cite{CMB-S4:2016ple} and CMB-HD at the level of $0.027$~\cite{Sehgal:2019ewc}.

The organization of the paper is as follows: In Sec.~\ref{sec:PDN}, we discuss the effective 3+1 pseudo-Dirac model (the derivation is in Appendix~\ref{app:A}) and the formalism for determination of $N_{\rm eff}$. In Sec.~\ref{sec:Neff}, we present our new bound on $\delta m_3^2$ and discuss future sensitivity. Finally, we conclude in Sec.~\ref{sec:conclusions}.

\section{Pseudo-Dirac neutrinos}\label{sec:PDN}

After the electroweak symmetry breaking and treating the SM as an effective field theory (EFT), both Majorana and Dirac mass terms are allowed. To write down the Dirac mass term, we have to introduce some new fermions or sterile neutrinos which do not feel any of the SM forces. While one can introduce only one sterile neutrino, we have opted for three sterile neutrinos $\nu'_{s_j}\,(j=1,2,3)$ such that each SM or so-called active neutrino has a Dirac partner. (In other words, we consider the SM as an EFT plus three sterile neutrinos.) In the lepton flavor basis where charged lepton
Yukawa is diagonal, we write the mass term as: $\overline{\Psi^{c}}{\cal M}\Psi$ for $\Psi=\left(\nu_{e},\nu_{\mu},\nu_{\tau},\nu_{s_{1}}',\nu_{s_{2}}',\nu_{s_{3}}'\right)$
with
\begin{eqnarray}
{\cal M} & = & \left(\begin{array}{cc}
m_{M} & m_{D}\\
m_{D}^{T} & m_{M}'
\end{array}\right), \label{eq:general_mass_matrix}
\end{eqnarray}
where $m_{D}$ is the Dirac mass term ($3\times 3$ complex matrix) which conserves the total lepton
number and $m_{M},m_{M}'$ are the Majorona mass terms ($3\times 3$ symmetric complex matrix) which break the total lepton number. Here $\nu_{e},\nu_{\mu},\nu_{\tau}$ are the SM neutrinos which feel the weak force and the subscripts indicate the charged leptons that they interact with.

The \emph{pseudo-Dirac} scenario will be defined as when the matrix entries in eq.~\eqref{eq:general_mass_matrix} satisfy
\begin{eqnarray}
\left|m_{M}\right|,\left|m_{M}'\right| & \ll & \left|m_{D}\right|,\label{eq:pseudo-Dirac}
\end{eqnarray}
which implies a small violation of the total lepton number.
Let us first consider the Dirac limit, $m_{M},m_{M}'\to 0$ where the total lepton number is conserved. In this
case, $m_{D}$ can be diagonalized by two unitary matrices $U$
and $V$ as follows
\begin{eqnarray}
\hat{m} & = & U^{T}m_{D}V=\textrm{diag}\left(m_{1},m_{2},m_{3}\right),\label{eq:neutrino_masses}
\end{eqnarray}
where $U$, in the charged lepton mass basis, gives rise to oscillations among active neutrinos and can be identified with the Pontecorvo–Maki–Nakagawa–Sakata (PMNS) matrix. On the other hand, $V$ which gives rise to oscillations among sterile neutrinos is of course not known.
Defining
\begin{eqnarray}
\mathcal{U} & = & \frac{1}{\sqrt{2}}\left(\begin{array}{cc}
U & iU\\
V & -iV
\end{array}\right),\label{eq:Dirac_limit}
\end{eqnarray}
the diagonalization of $M$ can be written as
\begin{equation}
\hat {\cal M} \equiv \mathcal{U}^{T} {\cal M} \mathcal{U} 
=\left(\begin{array}{cc}
\hat{m} & 0\\
0 & \hat{m}
\end{array}\right),
\end{equation}
where the fields in the mass basis are $\mathcal{U}^{\dagger}\Psi \equiv \left(\nu_{1},\nu_{2},\nu_{3},\nu'_{1},\nu'_{2},\nu'_{3}\right)^{T}$. Notice that $\nu_j$ and $\nu_j'$ have the same mass and represent a Dirac fermion, consisting of superposition of active and sterile neutrinos in equal proportion. Hence the  Dirac scenario is also known as the \emph{maximal mixing} scenario.

With $m_{M},m_{M}'\neq0$ but keeping the condition in eq.~\eqref{eq:pseudo-Dirac}, there will be modifications to eqs.~\eqref{eq:neutrino_masses} and \eqref{eq:Dirac_limit},
giving rise to three pairs of pseudo-Dirac states with masses ($j=1,2,3$)
\begin{eqnarray}
\hat {\cal M}_{j} & = & m_{j}-\delta m_{j},\quad \hat {\cal M}_{j+3}=m_{j}+\delta m_{j},\label{eq:mass_def}
\end{eqnarray}
where the pseudo-Dirac mass squared splitting is
\begin{equation}
    \delta m_j^2 \equiv \hat {\cal M}_{j+3}^2 - \hat {\cal M}_{j}^2 \simeq 4m_j \delta m_j.
\end{equation}
While the deviation from $U$ is of the order of $\delta U\sim m_{M}/m_{D},m'_{M}/m_{D}$, barring fine-tuning, the expected mass splitting is $|\delta m_j^{2}| \gtrsim 8 m_{j}^{2}|\delta U|$.
For example, taking $m_{j}\sim 0.1$ eV and considering $|\delta m_j^2| \lesssim 10^{-5}\,\textrm{eV}^{2}$ i.e. smaller than the solar mass squared splitting, we have $|\delta U| \lesssim 10^{-4}$. In general, for pseudo-Dirac scenario, the effect of $\delta U$ which controls the amplitude of oscillation probability will be dwarfed by the effect of new mass squared splitting which controls the oscillation frequency. 
Hence it is a good approximation to consider maximal mixing $\delta U = 0$ and study the effect of $\delta m_j^2 \neq 0$ as we will do in the rest of the work.
In other words, only when the effect of $\delta m_j^2 \neq 0$ is observed, one can start to consider the correction due to $\delta U \neq 0$.

\subsection{Effective 3+1 pseudo-Dirac model}

In principle, for the pseudo-Dirac scenario with three sterile neutrinos, one should consider a $3+3$ model. Nevertheless, assuming maximal active-sterile mixing and only one nonzero pseudo-Dirac mass splitting $\delta m_j^2$ at a time, one can work with an effective 3+1 Hamiltonian
\begin{eqnarray}
{\cal H}_{0,j}^{\rm eff} & \equiv & \frac{1}{2E}\left(\begin{array}{cc}
U\hat{m}^{2}U^\dagger & \epsilon_{j}\\
\epsilon_{j}^{\dagger} & m_{j}^{2}
\end{array}\right),\label{eq:eff31}
\end{eqnarray}
where $\epsilon_j \equiv -\frac{1}{2} \delta m_j^2 (U_{e j}, U_{\mu j}, U_{\tau j})^T$. The proof of the expression above is relegated to Appendix~\ref{app:A}. From eq.~\eqref{eq:eff31}, we see in this maximal mixing scenario, the transition to sterile state is controlled solely by $\delta m_j^2$ and as expected, in the purely Dirac scenario $\delta m_j^2 = 0$, one recovers the standard three-neutrino oscillation scenario. 

While we manage to obtain an effective 3+1 Hamiltonian for pseudo-Dirac scenario assuming one nonvanishing $\delta m_j^2$ at a time, note that it is qualitatively different from a 3+1 scenario (3 active plus one sterile neutrinos). 
To highlight the differences, let us consider a 3+1 scenario with a new mass eigenvalue with mass $m_4 \gg m_{1,2,3}$ and further assume that the new mixing elements involving the sterile state are small. Expanding in these small mixing elements, we obtain at leading order the same form as eq.~\eqref{eq:eff31} with $\epsilon_j \approx m_4^2 (U_{e 4}, U_{\mu 4}, U_{\tau 4})^T$. 
Under this approximation, we see that the 3+1 scenario depends on three new mixing elements $U_{\alpha 4} \,(\alpha = e, \mu, \tau)$ plus $m_4^2$ while for the pseudo-Dirac scenario, it depends only on one unknown $\delta m_j^2$ and the rest are mixing elements that been measured $U_{\alpha j}\, (\alpha = e, \mu, \tau; j=1,2,3)$. 
In the rest of the work, we will use the effective model of eq.~\eqref{eq:eff31} to derive new constraints on $\delta m_j^2$.

\subsection{Neutrino ensemble evolution in the early Universe}

In the hot plasma of the early Universe, neutrinos cannot in general be treated as free-streaming flavor states evolving only under vacuum oscillations and forward scattering. Instead, they form a dense, partially interacting ensemble that requires a quantum kinetic description. The central objects are the momentum $p$-dependent density matrices for neutrinos and antineutrinos at time $t$,
\begin{equation}
\varrho(p,t),\qquad \bar\varrho(p,t),
\end{equation}
which encode both the flavor populations (diagonal entries) and quantum coherences (off-diagonal entries). When we evolve in physical momentum $p$, cosmic expansion appears as a Liouville (redshift) term. 

We are interested in the cosmological temperatures from a few hundreds MeV down till tens of keV across the temperature when neutrinos decoupled from the thermal plasma. By considering the effective 3+1 pseudo-Dirac model, 
the evolution equation for the density matrix of neutrinos $[\varrho]_{ab},\,(a,b=e,\mu,\tau,s)$ reads~\cite{Sigl:1993ctk,Mirizzi:2012we,Saviano:2013ktj,deSalas:2016ztq,Gariazzo:2019gyi}
\begin{equation}
\big(\partial_t - H\,p\,\partial_p\big)\,\varrho(p,t)
= -\,i\big[\,
{\cal H}_{0,j}^{\rm eff}(p)
+V_{\rm mat}(t,p),\,\varrho(p,t)\,\big]
\;+\; C\!\left[\varrho (p,t),\bar\varrho (p,t)\right]
\label{eq:QKE-phys-p}
\end{equation}
and ignoring the asymmetries between number densities of particles and antiparticles (in the standard scenario where they are of the order of the observed baryon asymmetry, the effect is completely negligible), the corresponding equation for $\bar{\varrho}$ has the same form, with the usual replacement
$H_{0,j}^{\rm eff} \to \left(H_{0,j}^{\rm eff}\right)^{\ast}$.\footnote{In ref.~\cite{Sigl:1993ctk}, an equivalent convention is adopted for $\bar{\varrho}$ (with reversed flavor indices), which shifts the complex conjugation/sign between the Hamiltonian and the commutator term.}
Here ${\cal H}_{0,j}^{\rm eff}$ is given by eq.~\eqref{eq:eff31} with $E \simeq p$ and $H$ is the Hubble expansion rate. The first term on the right-hand side represents coherent flavor evolution, while $C[\varrho,\bar\varrho]$ encodes incoherent scatterings and repopulation. 

One can trade time $t$ with monotonically increasing scale factor $a$ through $H = (1/a)da/dt$ while the photon temperature $T$ will be determined using the continuity equation of the total plasma energy density $\rho_r$ (of $\gamma, e^{\pm}, \nu, \bar\nu$)
\begin{equation}
    \frac{d\rho_r}{dt} + 3 H (\rho_r + P ) = 0,
    \label{eq:continuity}
\end{equation}
where $P$ is the total pressure.

The matter potential $V_{\rm mat}$ arises from forward scattering of neutrinos with the medium. At a few hundreds MeV temperatures this includes charged-current contributions from $e^\pm, \mu^\pm$ and neutral-current contributions from the neutrino background itself. 
Unlike the familiar MSW potential proportional to net number density, here the relevant quantity is the finite-temperature energy density of the plasma. Following N\"{o}tzold and Raffelt, one finds, ignoring the asymmetries between particles and antiparticles~\cite{Notzold:1987ik}
\begin{equation}
V_{\rm mat}
= -\,\frac{8\sqrt{2}\,G_F\,p}{3}\!
\left(\frac{{\cal E}_\ell}{m_W^2}+\frac{{\cal E}_\nu}{m_Z^2}\right),
\end{equation}
where $G_F$ is the Fermi constant, $m_W$ and $m_Z$ are the $W$ and $Z$ boson masses, ${\cal E}_\ell=\mathrm{diag}(\rho_e,\rho_\mu,0,0)$ collects charged-lepton energy densities (summing over those of particles and antiparticles while those of tau leptons have been dropped since they are negligible for temperature $T \ll m_\tau$) and 
\begin{equation}
    {\cal E}_\nu = \frac{1}{2\pi^2}\int dp\, p^3 \, S_a (\varrho + \bar\varrho) S_a,
\end{equation}
where $S_a=\mathrm{diag}(1,1,1,0)$. (By definition, the sterile neutrinos do not have any SM interactions.)
Parametrically $V_{\rm mat}\propto G_F^2 T^5/\alpha$ where $\alpha$ is the fine-structure constant (one power from $p\!\sim\!T$ and four from the medium’s energy density), modifying the in-medium dispersion relation and the effective mixing with the sterile partner in the pseudo–Dirac scenario.

The collision term $C[\rho,\bar\rho]$ accounts for non-forward $2\leftrightarrow 2$ SM reactions,
\begin{equation}
\nu e^\pm \leftrightarrow \nu e^\pm,\qquad
e^+ e^- \leftrightarrow \nu\bar\nu,\qquad
\nu\nu(\bar\nu)\leftrightarrow \nu\nu(\bar\nu),
\end{equation}
including exact kinematics and Pauli blocking. (The ones involving $\mu^\pm$ are neglected since they play a role only at temperatures much higher than considered here.) These processes repopulate the distribution functions and damp quantum coherences, while oscillations redistribute flavors. 

The interplay of oscillations, matter effects, and collisions determines how efficiently the sterile partner of $\nu_j$ is populated prior to decoupling. 
Finally, the total neutrino energy density (including the sterile species)
\begin{equation}
    \rho_\nu = \frac{1}{2\pi^2}\int dp\, p^3 \,{\rm Tr}(\varrho + \bar\varrho),
\end{equation}
is encoded in $N_{\rm eff}$ which is defined as
\begin{equation}
    N_{\rm eff} \equiv \frac{8}{7} \left(\frac{11}{4}\right)^{4/3} \frac{\rho_\nu}{\rho_\gamma},
\end{equation}
where $\rho_\gamma$ is the photon energy density.

\section{$N_{\textrm{eff}}$ determination} \label{sec:Neff}

\begin{figure}
\begin{centering}
\includegraphics[scale=0.9]{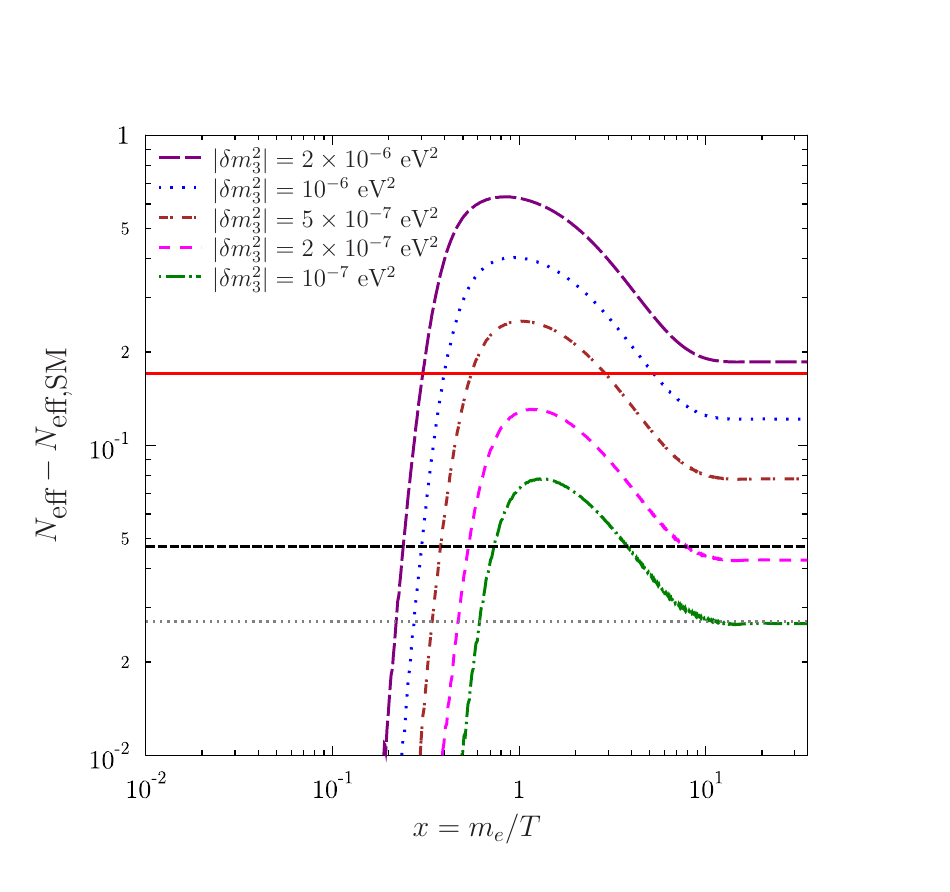}
\par\end{centering}
\caption{$N_{\rm eff} - N_{\rm eff,SM}$ as a function of $x = m_e/T$ for various values of $|\delta m_3^2|$. The red solid line denotes at 2$\sigma$ level, the latest bound from ACT + Planck + BAO~\cite{Planck:2018vyg,AtacamaCosmologyTelescope:2025nti}, the dashed black line denotes the sensitivity of CMB-S4~\cite{CMB-S4:2016ple} and the dotted black line denotes the sensitivity of CMB-HD~\cite{Sehgal:2019ewc}.
	\label{fig:main_result}
}
\end{figure}

To derive the constraint on $N_{\rm eff}$, we implement our effective 3+1 pseudo-Dirac model in the public code \texttt{FortEPiaNO}~\cite{Gariazzo:2019gyi} which solves eqs.~\eqref{eq:QKE-phys-p} and \eqref{eq:continuity} for each momentum mode $p$ of neutrinos and antineutrinos i.e. by discretizing $p$. For the PMNS matrix $U$ and the standard mass squared splitting $m_2^2 - m_1^2$ and $m_3^2 - m_1^2$, we use the best-fit values given by ref.~\cite{Esteban:2024eli}. Following refs.~\cite{Gariazzo:2019gyi,Bennett:2020zkv}, the CP phase in $U$ has been set to zero and therefore, neutrinos and antineutrinos behave identically. For our study, the effect of the CP phase is expected to be negligible.

Firstly, we have verified that the mass ordering of neutrinos (i.e. for normal ordering $m_1 < m_2 < m_3$ or inverse ordering $m_3 < m_1 < m_2$) gives a negligible difference in $N_{\rm eff}$ similar to the result for the SM scenario~\cite{Bennett:2020zkv}. 
Secondly, we have checked that the sign of $\delta m_j^2$ plays no role when the active-sterile mixing is almost maximal.\footnote{In principle, resonant conversion could occur for $\delta m_j^2 < 0$~\cite{Dolgov:2003sg}. Nevertheless, it can be shown that for pseudo-Dirac scenario with $|\delta m_j^2| \gtrsim 8m_j^2 |\delta U|$, this only happens when $\delta m_j^2 \lesssim 10^{-23}\,{\rm eV}^2$.}
Thirdly, we have checked that the bounds we obtain have negligible differences among $j=1,2,3$. In the view of stronger bounds on $|\delta m_{1,2}^2| \lesssim 10^{-11}\,{\rm eV}^2$ from solar neutrino data~\cite{Cirelli:2004cz,deGouvea:2009fp,Anamiati:2017rxw,Ansarifard:2022kvy}, we will present our result below only for $|\delta m_3^2| \neq 0$.

Our main result is shown in Fig.~\ref{fig:main_result} where for different values of $|\delta m_3^2|$, the corresponding $N_{\rm eff} - N_{\rm eff,SM}$ is shown as a function of $x = m_e/T$ where $m_e$ is the electron mass
and $N_{\rm eff,SM} = 3.0440$ is the SM value. The latest CMB measurements from the Atacama Cosmology Telescope (ACT) in combination with Planck and BAO measurements give $N_{\rm eff} - N_{\rm eff,SM}<0.17$ at 95\% CL~\cite{Planck:2018vyg,AtacamaCosmologyTelescope:2025nti} which translates to $|\delta m_3^2| < 2\times 10^{-6}\,{\rm eV}^2$, which is about an order of magnitude stronger than the constraint from neutrino oscillation data~\cite{Anamiati:2017rxw}. We further show that the future ground-based telescopes CMB-S4~\cite{CMB-S4:2016ple} and CMB-HD~\cite{Sehgal:2019ewc} can probe down to $|\delta m_3^2| \sim 10^{-7}\,{\rm eV}^2$.

\section{Conclusions} \label{sec:conclusions}

In this work, we have derived an effective 3+1 Hamiltonian for pseudo-Dirac scenario assuming maximal active-sterile mixing and one nonzero pseudo-Dirac mass splitting $\delta m^2_j$. Implementing this in the state-of-the-art public code \texttt{FortEPiaNO}~\cite{Gariazzo:2019gyi}, we have shown that the current CMB data is already constraining $|\delta m_3^2| < 2\times 10^{-6}\,{\rm eV}^2$, about an order of magnitude stronger than using the neutrino oscillation data. Future CMB measurements will be able to increase the sensitivy by an order of magnitude down to $|\delta m_3^2| \sim 10^{-7}\,{\rm eV}^2$. 
One future direction is to combine this study with the BBN which can provide further collaborative constraint~\cite{Dolgov:2003sg,Cirelli:2004cz}.
The other future direction is to consider a nonstandard scenario where the lepton asymmetry can be much larger than expected (in comparison to the baryon asymmetry), leading to a difference in matter potential experienced by neutrinos and antineutrinos~\cite{Dolgov:2003sg,Chu:2006ua,Mirizzi:2012we,Saviano:2013ktj}.

\section*{Acknowledgments}
 C.S.F. acknowledges the support from the Gordon Godfrey Bequest during his one-month visit to University of New South Wales in August 2023 where this project was initiated. He also acknowledges the support by Fundação de Amparo à Pesquisa do Estado de São Paulo (FAPESP) Contracts No. 2019/11197-6 and 2022/00404-3 and Conselho Nacional de Desenvolvimento Científico e Tecnológico (CNPq) under Contracts No. 407149/2021-0 and No. 304917/2023-0. 
 Y.P. acknowledges the support by FAPESP Contract No.  2023/10734-3 and 2023/01467-1, and the support by the DFG Collaborative Research Institution Neutrinos and Dark
Matter in Astro- and Particle Physics (SFB 1258).
 The authors are grateful to Yvonne Y. Y. Wong for discussion and Stefano Gariazzo for the help with \texttt{FortEPiaNO}.
 The authors acknowledge the Center for Theoretical Underground Physics and Related Areas (CETUP* 2025) and the Institute for Underground Science at Sanford Underground Research Facility (SURF) for hospitality and for providing a conducive environment where this work was finalized.

\appendix

\section{Derivation of effective pseudo-Dirac Hamiltonian}\label{app:A}

For pseudo-Dirac neutrinos, a useful parametrization for unitary matrix
${\cal U}$ which relates neutrinos in the flavor basis (charged lepton Yukawa is diagonal) and the mass
basis $\Psi={\cal U}\hat{\Psi}$ with $\Psi=\left(\nu_{e},\nu_{\mu},\nu_{\tau},\nu_{s_{1}}',\nu_{s_{2}}',\nu_{s_{3}}'\right)^{T}$
and $\hat{\Psi}=\left(\nu_{1},\nu_{2},\nu_{3},\nu_{1}',\nu_{2}',\nu_{3}'\right)^{T}$,
is given by
\begin{equation}
{\cal U}=U_{\textrm{new}}UY,
\end{equation}
where
\begin{eqnarray}
U_{\textrm{new}} & \equiv & R_{56}R_{46}R_{36}R_{26}R_{16}R_{45}R_{35}R_{25}R_{15}R_{34}R_{24}R_{14},\label{eq:U_new}\\
U & \equiv & R_{23}R_{13}R_{12},\label{eq:U}\\
Y & \equiv & \frac{1}{\sqrt{2}}\left(\begin{array}{cc}
I_{3\times3} & iI_{3\times3}\\
I_{3\times3} & -iI_{3\times3}
\end{array}\right).\label{eq:Y}
\end{eqnarray}
In the above, $I_{3\times3}$ denotes the $3\times3$ identity matrix
and $R_{ij}$ is the complex rotation matrix in the $ij$-plane which
can be constructed starting from a $6\times6$ identity matrix $I$
and then replacing $I_{ii}$ and $I_{jj}$ by $\cos\theta_{ij}$,
$I_{ij}$ by $e^{-i\phi_{ij}}\sin\theta_{ij}$ and $I_{ji}$ by $e^{i\phi_{ij}}\sin\theta_{ij}$.
The Dirac limit i.e. \emph{maximal mixing} is obtained by setting
all the angles besides $\theta_{45}$, $\theta_{46}$, $\theta_{56}$
in eq. (\ref{eq:U_new}) to zero giving 
\begin{eqnarray}
{\cal U} & = & \frac{1}{\sqrt{2}}\left(\begin{array}{cc}
U & iU\\
V & -iV
\end{array}\right)=\frac{1}{\sqrt{2}}\left(\begin{array}{cc}
U & 0_{3\times3}\\
0_{3\times3} & V
\end{array}\right)Y,\label{eq:maximal_mixing}
\end{eqnarray}
where $U$ and $V$ are both unitary matrices. While $U_{0}$ is identified
with the PMNS matrix, $V$ specified by the three angles $\theta_{45}$,
$\theta_{46}$, $\theta_{56}$ and the corresponding phases give rise
to the mixing among the sterile neutrinos and are not measurable through
the SM interactions. 

Since the effect of deviation from the maximal mixing is much smaller
than the effect of the new mass splitting, we will consider the maximal mixing
as in eq.~\eqref{eq:maximal_mixing} while keeping $\delta m_{j}^{2}\neq0$.
Next, we will prove that $V$ will not play a role. Carrying out
a rotation in the flavor space of sterile neutrinos as follows
\begin{eqnarray}
\left(\nu_{s_{1}}',\nu_{s_{2}}',\nu_{s_{3}}'\right)^{T} & \to & V\left(\nu_{s_{1}}',\nu_{s_{2}}',\nu_{s_{3}}'\right)^{T},
\end{eqnarray}
eq.~\eqref{eq:maximal_mixing} becomes
\begin{eqnarray}
{\cal U} & \to & \frac{1}{\sqrt{2}}\left(\begin{array}{cc}
U & 0_{3\times3}\\
0_{3\times3} & I_{3\times3}
\end{array}\right)Y,\label{eq:V_identity_basis}
\end{eqnarray}
where $V$ has disappeared. This can always be done provided there
is no interaction that allows to distinguish sterile flavors as it is assumed in this work. 

The free Hamiltonian in the neutrino flavor basis is given by
\begin{eqnarray}
{\cal H}_{0} & = & {\cal U}\Delta{\cal U}^{\dagger},\label{eq:free_Hamiltonian}
\end{eqnarray}
where 
\begin{eqnarray}
\Delta & \equiv & \frac{1}{2E}{\rm diag}\left(\hat{{\cal M}}_{1}^{2},\hat{{\cal M}}_{2}^{2},\hat{{\cal M}}_{3}^{2},\hat{{\cal M}}_{4}^{2},\hat{{\cal M}}_{5}^{2},\hat{{\cal M}}_{6}^{2}\right).
\end{eqnarray}
By allowing only one nonzero $\delta m_{j}^{2}$ at a time and using
eq. (\ref{eq:V_identity_basis}) in eq. (\ref{eq:free_Hamiltonian}),
we obtain an effective $3+1$ Hamiltonian given by
\begin{eqnarray}
{\cal H}_{0,j}^{{\rm eff}} & \equiv & \frac{1}{2E}\left(\begin{array}{cc}
U\hat{m}^{2}U^\dagger & \epsilon_{j}\\
\epsilon_{j}^{\dagger} & m_{j}^{2}
\end{array}\right),
\end{eqnarray}
where $\hat{m}^{2}={\rm diag}\left(m_{1}^{2},m_{2}^{2},m_{3}^{2}\right)$
and $\epsilon_{j}\equiv-\frac{1}{2}\delta m_{j}^{2}\left(U_{e j},U_{\mu j},U_{\tau j}\right)^T$.
For numerical implementation, it can be more convenient to rewrite the above as
\begin{eqnarray}
{\cal H}_{0,j}^{{\rm eff}} & = & \frac{1}{2E}U_{4}\left(\begin{array}{cccc}
m_{1}^{2} & 0 & 0 & -\frac{1}{2}\delta m_{j}^{2}\delta_{1j}\\
0 & m_{2}^{2} & 0 & -\frac{1}{2}\delta m_{j}^{2}\delta_{2j}\\
0 & 0 & m_{3}^{2} & -\frac{1}{2}\delta m_{j}^{2}\delta_{3j}\\
-\frac{1}{2}\delta m_{j}^{2}\delta_{1j} & -\frac{1}{2}\delta m_{j}^{2}\delta_{2j} & -\frac{1}{2}\delta m_{j}^{2}\delta_{3j} & m_{j}^{2}
\end{array}\right)U_{4}^{\dagger},
\end{eqnarray}
where
\begin{eqnarray}
U_{4} & \equiv & \left(\begin{array}{cc}
U & 0_{3\times1}\\
0_{1\times3} & 1
\end{array}\right).
\end{eqnarray}

\bibliography{PD_refs}

@article{Valle:1982yw,
    author = "Valle, J. W. F.",
    title = "{Neutrinoless Double Beta Decay With Quasi Dirac Neutrinos}",
    reportNumber = "PRINT-82-0694 (SYRACUSE)",
    doi = "10.1103/PhysRevD.27.1672",
    journal = "Phys. Rev. D",
    volume = "27",
    pages = "1672--1674",
    year = "1983"
}

@article{Wolfenstein:1981kw,
    author = "Wolfenstein, Lincoln",
    title = "{Different Varieties of Massive Dirac Neutrinos}",
    reportNumber = "COO-3066-164",
    doi = "10.1016/0550-3213(81)90096-1",
    journal = "Nucl. Phys. B",
    volume = "186",
    pages = "147--152",
    year = "1981"
}

@article{Petcov:1982ya,
    author = "Petcov, S. T.",
    title = "{On Pseudodirac Neutrinos, Neutrino Oscillations and Neutrinoless Double beta Decay}",
    doi = "10.1016/0370-2693(82)91246-1",
    journal = "Phys. Lett. B",
    volume = "110",
    pages = "245--249",
    year = "1982"
}

@article{Anamiati:2017rxw,
    author = "Anamiati, G. and Fonseca, R. M. and Hirsch, M.",
    title = "{Quasi Dirac neutrino oscillations}",
    eprint = "1710.06249",
    archivePrefix = "arXiv",
    primaryClass = "hep-ph",
    reportNumber = "IFIC-17-45",
    doi = "10.1103/PhysRevD.97.095008",
    journal = "Phys. Rev. D",
    volume = "97",
    number = "9",
    pages = "095008",
    year = "2018"
}

@article{Anamiati:2019maf,
    author = "Anamiati, G. and De Romeri, V. and Hirsch, M. and Ternes, C. A. and T{\'o}rtola, M.",
    title = "{Quasi-Dirac neutrino oscillations at DUNE and JUNO}",
    eprint = "1907.00980",
    archivePrefix = "arXiv",
    primaryClass = "hep-ph",
    doi = "10.1103/PhysRevD.100.035032",
    journal = "Phys. Rev. D",
    volume = "100",
    number = "3",
    pages = "035032",
    year = "2019"
}

@article{Fong:2020smz,
    author = "Fong, C. S. and Gregoire, T. and Tonero, A.",
    title = "{Testing quasi-Dirac leptogenesis through neutrino oscillations}",
    eprint = "2007.09158",
    archivePrefix = "arXiv",
    primaryClass = "hep-ph",
    doi = "10.1016/j.physletb.2021.136175",
    journal = "Phys. Lett. B",
    volume = "816",
    pages = "136175",
    year = "2021"
}

@article{Cirelli:2004cz,
    author = "Cirelli, Marco and Marandella, Guido and Strumia, Alessandro and Vissani, Francesco",
    title = "{Probing oscillations into sterile neutrinos with cosmology, astrophysics and experiments}",
    eprint = "hep-ph/0403158",
    archivePrefix = "arXiv",
    reportNumber = "IFUP-TH-2004-2",
    doi = "10.1016/j.nuclphysb.2004.11.056",
    journal = "Nucl. Phys. B",
    volume = "708",
    pages = "215--267",
    year = "2005"
}

@article{deGouvea:2009fp,
    author = "de Gouvea, Andre and Huang, Wei-Chih and Jenkins, James",
    title = "{Pseudo-Dirac Neutrinos in the New Standard Model}",
    eprint = "0906.1611",
    archivePrefix = "arXiv",
    primaryClass = "hep-ph",
    reportNumber = "LA-UR-09-03593, NUHEP-TH-09-08",
    doi = "10.1103/PhysRevD.80.073007",
    journal = "Phys. Rev. D",
    volume = "80",
    pages = "073007",
    year = "2009"
}

@article{Ansarifard:2022kvy,
    author = "Ansarifard, Saeed and Farzan, Yasaman",
    title = "{Revisiting pseudo-Dirac neutrino scenario after recent solar neutrino data}",
    eprint = "2211.09105",
    archivePrefix = "arXiv",
    primaryClass = "hep-ph",
    doi = "10.1103/PhysRevD.107.075029",
    journal = "Phys. Rev. D",
    volume = "107",
    number = "7",
    pages = "075029",
    year = "2023"
}

@article{Keranen:2003xd,
    author = "Keranen, P. and Maalampi, J. and Myyrylainen, M. and Riittinen, J.",
    title = "{Effects of sterile neutrinos on the ultrahigh-energy cosmic neutrino flux}",
    eprint = "hep-ph/0307041",
    archivePrefix = "arXiv",
    reportNumber = "JYFL-HE-8-2003, NORDITA-2003-50HEP",
    doi = "10.1016/j.physletb.2003.09.006",
    journal = "Phys. Lett. B",
    volume = "574",
    pages = "162--168",
    year = "2003"
}

@article{Beacom:2003eu,
    author = "Beacom, John F. and Bell, Nicole F. and Hooper, Dan and Learned, John G. and Pakvasa, Sandip and Weiler, Thomas J.",
    title = "{PseudoDirac Neutrinos: A Challenge for Neutrino Telescopes}",
    eprint = "hep-ph/0307151",
    archivePrefix = "arXiv",
    reportNumber = "FERMILAB-PUB-03-201-A, MADPH-03-1337",
    doi = "10.1103/PhysRevLett.92.011101",
    journal = "Phys. Rev. Lett.",
    volume = "92",
    pages = "011101",
    year = "2004"
}

@article{Esmaili:2009fk,
    author = "Esmaili, Arman",
    title = "{Pseudo-Dirac Neutrino Scenario: Cosmic Neutrinos at Neutrino Telescopes}",
    eprint = "0909.5410",
    archivePrefix = "arXiv",
    primaryClass = "hep-ph",
    reportNumber = "IPM-P-2009-039",
    doi = "10.1103/PhysRevD.81.013006",
    journal = "Phys. Rev. D",
    volume = "81",
    pages = "013006",
    year = "2010"
}

@article{Esmaili:2012ac,
    author = "Esmaili, Arman and Farzan, Yasaman",
    title = "{Implications of the Pseudo-Dirac Scenario for Ultra High Energy Neutrinos from GRBs}",
    eprint = "1208.6012",
    archivePrefix = "arXiv",
    primaryClass = "hep-ph",
    doi = "10.1088/1475-7516/2012/12/014",
    journal = "JCAP",
    volume = "12",
    pages = "014",
    year = "2012"
}

@article{Joshipura:2013yba,
    author = "Joshipura, Anjan S. and Mohanty, Subhendra and Pakvasa, Sandip",
    title = "{Pseudo-Dirac neutrinos via a mirror world and depletion of ultrahigh energy neutrinos}",
    eprint = "1307.5712",
    archivePrefix = "arXiv",
    primaryClass = "hep-ph",
    doi = "10.1103/PhysRevD.89.033003",
    journal = "Phys. Rev. D",
    volume = "89",
    number = "3",
    pages = "033003",
    year = "2014"
}

@article{Brdar:2018tce,
    author = "Brdar, Vedran and Hansen, Rasmus S. L.",
    title = "{IceCube Flavor Ratios with Identified Astrophysical Sources: Towards Improving New Physics Testability}",
    eprint = "1812.05541",
    archivePrefix = "arXiv",
    primaryClass = "hep-ph",
    doi = "10.1088/1475-7516/2019/02/023",
    journal = "JCAP",
    volume = "02",
    pages = "023",
    year = "2019"
}

@article{DeGouvea:2020ang,
    author = "De Gouv{\^e}a, Andr{\'e} and Martinez-Soler, Ivan and Perez-Gonzalez, Yuber F. and Sen, Manibrata",
    title = "{Fundamental physics with the diffuse supernova background neutrinos}",
    eprint = "2007.13748",
    archivePrefix = "arXiv",
    primaryClass = "hep-ph",
    reportNumber = "NUHEP-TH/20-08, FERMILAB-PUB-20-353-T",
    doi = "10.1103/PhysRevD.102.123012",
    journal = "Phys. Rev. D",
    volume = "102",
    pages = "123012",
    year = "2020"
}

@article{Martinez-Soler:2021unz,
    author = "Martinez-Soler, Ivan and Perez-Gonzalez, Yuber F. and Sen, Manibrata",
    title = "{Signs of pseudo-Dirac neutrinos in SN1987A data}",
    eprint = "2105.12736",
    archivePrefix = "arXiv",
    primaryClass = "hep-ph",
    reportNumber = "FERMILAB-PUB-21-225-T, NUHEP-TH/21-05, N3AS-21-009",
    doi = "10.1103/PhysRevD.105.095019",
    journal = "Phys. Rev. D",
    volume = "105",
    number = "9",
    pages = "095019",
    year = "2022"
}

@article{Rink:2022nvw,
    author = "Rink, Thomas and Sen, Manibrata",
    title = "{Constraints on pseudo-Dirac neutrinos using high-energy neutrinos from NGC 1068}",
    eprint = "2211.16520",
    archivePrefix = "arXiv",
    primaryClass = "hep-ph",
    doi = "10.1016/j.physletb.2024.138558",
    journal = "Phys. Lett. B",
    volume = "851",
    pages = "138558",
    year = "2024"
}

@article{Carloni:2022cqz,
    author = "Carloni, Kiara and Mart{\'\i}nez-Soler, Ivan and Arguelles, Carlos A. and Babu, K. S. and Dev, P. S. Bhupal",
    title = "{Probing pseudo-Dirac neutrinos with astrophysical sources at IceCube}",
    eprint = "2212.00737",
    archivePrefix = "arXiv",
    primaryClass = "astro-ph.HE",
    doi = "10.1103/PhysRevD.109.L051702",
    journal = "Phys. Rev. D",
    volume = "109",
    pages = "L051702",
    year = "2024"
}

@article{Kirilova:1997sv,
    author = "Kirilova, D. P. and Chizhov, M. V.",
    title = "{Cosmological nucleosynthesis and active sterile neutrino oscillations with small mass differences: The Nonresonant case}",
    eprint = "hep-ph/9707282",
    archivePrefix = "arXiv",
    reportNumber = "TAC-1997-019",
    doi = "10.1103/PhysRevD.58.073004",
    journal = "Phys. Rev. D",
    volume = "58",
    pages = "073004",
    year = "1998"
}

@article{Kirilova:1999xj,
    author = "Kirilova, D. P. and Chizhov, M. V.",
    title = "{Cosmological nucleosynthesis and active sterile neutrino oscillations with small mass differences: The Resonant case}",
    eprint = "hep-ph/9909408",
    archivePrefix = "arXiv",
    reportNumber = "IC-99-127",
    doi = "10.1016/S0550-3213(00)00541-1",
    journal = "Nucl. Phys. B",
    volume = "591",
    pages = "457--468",
    year = "2000"
}

@article{Kirilova:2006wh,
    author = "Kirilova, Daniela P. and Panayotova, Mariana P.",
    title = "{Relaxed constraints on neutrino oscillation parameters}",
    eprint = "astro-ph/0608103",
    archivePrefix = "arXiv",
    doi = "10.1088/1475-7516/2006/12/014",
    journal = "JCAP",
    volume = "12",
    pages = "014",
    year = "2006"
}

@article{Dolgov:2003sg,
    author = "Dolgov, A. D. and Villante, F. L.",
    title = "{BBN bounds on active sterile neutrino mixing}",
    eprint = "hep-ph/0308083",
    archivePrefix = "arXiv",
    doi = "10.1016/j.nuclphysb.2003.11.031",
    journal = "Nucl. Phys. B",
    volume = "679",
    pages = "261--298",
    year = "2004"
}

@article{Planck:2018vyg,
    author = "Aghanim, N. and others",
    collaboration = "Planck",
    title = "{Planck 2018 results. VI. Cosmological parameters}",
    eprint = "1807.06209",
    archivePrefix = "arXiv",
    primaryClass = "astro-ph.CO",
    doi = "10.1051/0004-6361/201833910",
    journal = "Astron. Astrophys.",
    volume = "641",
    pages = "A6",
    year = "2020",
    note = "[Erratum: Astron.Astrophys. 652, C4 (2021)]"
}

@article{Bennett:2020zkv,
    author = "Bennett, Jack J. and Buldgen, Gilles and De Salas, Pablo F. and Drewes, Marco and Gariazzo, Stefano and Pastor, Sergio and Wong, Yvonne Y. Y.",
    title = "{Towards a precision calculation of $N_{\rm eff}$ in the Standard Model II: Neutrino decoupling in the presence of flavour oscillations and finite-temperature QED}",
    eprint = "2012.02726",
    archivePrefix = "arXiv",
    primaryClass = "hep-ph",
    reportNumber = "CPPC-2020-10",
    doi = "10.1088/1475-7516/2021/04/073",
    journal = "JCAP",
    volume = "04",
    pages = "073",
    year = "2021"
}

@book{CMB-S4:2016ple,
    author = "Abazajian, Kevork N. and others",
    collaboration = "CMB-S4",
    title = "{CMB-S4 Science Book, First Edition}",
    eprint = "1610.02743",
    archivePrefix = "arXiv",
    primaryClass = "astro-ph.CO",
    reportNumber = "FERMILAB-FN-1024-A-AE",
    doi = "10.2172/1352047",
    month = "10",
    year = "2016"
}

@article{Gariazzo:2019gyi,
    author = "Gariazzo, S. and de Salas, P. F. and Pastor, S.",
    title = "{Thermalisation of sterile neutrinos in the early Universe in the 3+1 scheme with full mixing matrix}",
    eprint = "1905.11290",
    archivePrefix = "arXiv",
    primaryClass = "astro-ph.CO",
    doi = "10.1088/1475-7516/2019/07/014",
    journal = "JCAP",
    volume = "07",
    pages = "014",
    year = "2019"
}

@article{Sehgal:2019ewc,
    author = "Sehgal, Neelima and others",
    title = "{CMB-HD: An Ultra-Deep, High-Resolution Millimeter-Wave Survey Over Half the Sky}",
    eprint = "1906.10134",
    archivePrefix = "arXiv",
    primaryClass = "astro-ph.CO",
    journal = "Bull. Am. Astron. Soc.",
    volume = "51",
    number = "7",
    pages = "1--23",
    year = "2019"
}

@article{Fong:2024msb,
    author = "Fong, Chee Sheng and Porto, Yago",
    title = "{Constraining the pseudo-Dirac nature of neutrinos using astrophysical neutrino flavor data}",
    eprint = "2406.15566",
    archivePrefix = "arXiv",
    primaryClass = "hep-ph",
    doi = "10.1103/yc27-61w5",
    journal = "Phys. Rev. D",
    volume = "112",
    number = "6",
    pages = "063001",
    year = "2025"
}

@article{Dev:2024yrg,
    author = "Dev, P. S. Bhupal and Machado, Pedro A. N. and Martinez-Soler, Ivan",
    title = "{Pseudo-Dirac neutrinos and relic neutrino matter effect on the high-energy neutrino flavor composition}",
    eprint = "2406.18507",
    archivePrefix = "arXiv",
    primaryClass = "hep-ph",
    reportNumber = "CETUP-2023-022, FERMILAB-PUB-24-0317-T, IPPP/24/35",
    doi = "10.1016/j.physletb.2025.139306",
    journal = "Phys. Lett. B",
    volume = "862",
    pages = "139306",
    year = "2025"
}

@article{Esteban:2024eli,
    author = "Esteban, Ivan and Gonzalez-Garcia, M. C. and Maltoni, Michele and Martinez-Soler, Ivan and Pinheiro, Jo{\~a}o Paulo and Schwetz, Thomas",
    title = "{NuFit-6.0: updated global analysis of three-flavor neutrino oscillations}",
    eprint = "2410.05380",
    archivePrefix = "arXiv",
    primaryClass = "hep-ph",
    reportNumber = "IFT-UAM/CSIC-24-140, YITP-SB-2024-24, IPPP/24/64, IPPP/24/64, IFT-UAM/CSIC-24-140, YITP-SB-2024-24",
    doi = "10.1007/JHEP12(2024)216",
    journal = "JHEP",
    volume = "12",
    pages = "216",
    year = "2024"
}

@article{Schoneberg:2024ifp,
    author = {Sch{\"o}neberg, Nils},
    title = "{The 2024 BBN baryon abundance update}",
    eprint = "2401.15054",
    archivePrefix = "arXiv",
    primaryClass = "astro-ph.CO",
    doi = "10.1088/1475-7516/2024/06/006",
    journal = "JCAP",
    volume = "06",
    pages = "006",
    year = "2024"
}

@article{Notzold:1987ik,
    author = {N{\"o}tzold, Dirk and Raffelt, Georg},
    title = "{Neutrino dispersion at finite temperature and density}",
    reportNumber = "MPI-PAE/PTh-87/87",
    doi = "10.1016/0550-3213(88)90113-7",
    journal = "Nucl. Phys. B",
    volume = "307",
    pages = "924--936",
    year = "1988"
}

@article{Sigl:1993ctk,
    author = "Sigl, G. and Raffelt, G.",
    title = "{General kinetic description of relativistic mixed neutrinos}",
    reportNumber = "MPI-PH-92-112",
    doi = "10.1016/0550-3213(93)90175-O",
    journal = "Nucl. Phys. B",
    volume = "406",
    pages = "423--451",
    year = "1993"
}

@article{Mirizzi:2012we,
    author = "Mirizzi, Alessandro and Saviano, Ninetta and Miele, Gennaro and Serpico, Pasquale Dario",
    title = "{Light sterile neutrino production in the early universe with dynamical neutrino asymmetries}",
    eprint = "1206.1046",
    archivePrefix = "arXiv",
    primaryClass = "hep-ph",
    reportNumber = "LAPTH-026-12",
    doi = "10.1103/PhysRevD.86.053009",
    journal = "Phys. Rev. D",
    volume = "86",
    pages = "053009",
    year = "2012"
}

@article{Saviano:2013ktj,
    author = "Saviano, Ninetta and Mirizzi, Alessandro and Pisanti, Ofelia and Serpico, Pasquale Dario and Mangano, Gianpiero and Miele, Gennaro",
    title = "{Multi-momentum and multi-flavour active-sterile neutrino oscillations in the early universe: role of neutrino asymmetries and effects on nucleosynthesis}",
    eprint = "1302.1200",
    archivePrefix = "arXiv",
    primaryClass = "astro-ph.CO",
    reportNumber = "DF-2013-3, LAPTH-004-13",
    doi = "10.1103/PhysRevD.87.073006",
    journal = "Phys. Rev. D",
    volume = "87",
    pages = "073006",
    year = "2013"
}

@article{deSalas:2016ztq,
    author = "de Salas, Pablo F. and Pastor, Sergio",
    title = "{Relic neutrino decoupling with flavour oscillations revisited}",
    eprint = "1606.06986",
    archivePrefix = "arXiv",
    primaryClass = "hep-ph",
    reportNumber = "IFIC-16-10, TTK-16-23",
    doi = "10.1088/1475-7516/2016/07/051",
    journal = "JCAP",
    volume = "07",
    pages = "051",
    year = "2016"
}

@article{Chu:2006ua,
    author = "Chu, Yi-Zen and Cirelli, Marco",
    title = "{Sterile neutrinos, lepton asymmetries, primordial elements: How much of each?}",
    eprint = "astro-ph/0608206",
    archivePrefix = "arXiv",
    doi = "10.1103/PhysRevD.74.085015",
    journal = "Phys. Rev. D",
    volume = "74",
    pages = "085015",
    year = "2006"
}

@article{AtacamaCosmologyTelescope:2025nti,
    author = "Calabrese, Erminia and others",
    collaboration = "Atacama Cosmology Telescope",
    title = "{The Atacama Cosmology Telescope: DR6 constraints on extended cosmological models}",
    eprint = "2503.14454",
    archivePrefix = "arXiv",
    primaryClass = "astro-ph.CO",
    reportNumber = "FERMILAB-PUB-25-0157-PPD",
    doi = "10.1088/1475-7516/2025/11/063",
    journal = "JCAP",
    volume = "11",
    pages = "063",
    year = "2025"
}

@article{Carloni:2025dhv,
    author = {Carloni, Kiara and Porto, Yago and Arg{\"u}elles, Carlos A. and Dev, P. S. Bhupal and Jana, Sudip},
    title = "{Signatures of quasi-Dirac neutrinos in diffuse high-energy astrophysical neutrino data}",
    eprint = "2503.19960",
    archivePrefix = "arXiv",
    primaryClass = "hep-ph",
    month = "3",
    year = "2025"
}

\end{document}